\documentclass{aa}

\input psfig.tex

\usepackage{graphicx}
\usepackage{natbib}
\usepackage{array}
\bibpunct{(}{)}{;}{a}{}{,}

\usepackage{rotating}    

\input psfig.tex
\usepackage{latexsym}
\usepackage{natbib}
\usepackage{amssymb}
\usepackage{amsmath}

\usepackage{graphicx}
\usepackage{graphics}
\usepackage{fancyhdr}
\usepackage{morefloats}

\begin{document}

\title{Simulating the formation and evolution of galaxies: \\
 Multi-phase description of the interstellar medium, star formation, and energy feedback}
\subtitle{}
     \author{ Emiliano Merlin \inst{1} \&
           Cesare Chiosi \inst{1}
            }
     \offprints{E. Merlin }
     \institute{$^1 $ Department of Astronomy, University of Padova,
                Vicolo dell'Osservatorio 3, 35122 Padova, Italy  \\
                \email{emiliano.merlin@unipd.it; cesare.chiosi@unipd.it }
      }
     \date{Received: May 2007; Accepted:  }


\abstract {Modelling  the gaseous component of the interstellar
medium (ISM) by Smoothed Particles Hydrodynamics in N-Body
simulations (NB-TSPH) is still very crude when compared to the
complex real situation. In the real ISM, many different and almost
physically decoupled components (phases) coexist for long periods of
time, and since they spread over wide ranges of density and
temperature, they cannot be correctly represented by a unique
continuous fluid. This would certainly influence star formation
which is thought to take place in clumps of cold, dense, molecular
clouds, embedded in a warmer, neutral medium, that are almost freely
moving throughout the tenuous hot ISM. Therefore, assuming that star
formation is simply related to the gas content without specifying
the component in which this is both observed and expected to occur,
may not be physically sounded.} {In this paper we seek to improve
upon this aspect by considering a multi-phase representation of the
ISM in NB-TSPH simulations of galaxy formation and evolution with
particular attention to the case of early-type galaxies.} {Cold gas
clouds are described by the so-called \textit{sticky particles}
algorithm. They can freely move throughout the hot ISM medium; stars
form within these clouds and the mass exchange among the three
baryonic phases (hot gas, cold clouds, stars) is governed by
radiative and Compton cooling and energy feedback by supernova (SN)
explosions, stellar winds, and UV radiation. We also consider
thermal conduction, cloud-cloud collisions, and chemical
enrichment.} {Our model agrees with and improves upon previous
studies on the same subject. The results for the star formation rate
are very promising and agree with recent observational data on
early-type galaxies.} {These models lend further support to the
revised monolithic scheme of galaxy formation, which has recently
been also strengthened by high redshift data leading to the
so-called \textit{downsizing} and \textit{top-down}
scenarios.}

\titlerunning{Multi-phase description  of the ISM in early-type galaxies}
\authorrunning{E. Merlin \& C. Chiosi }
\maketitle


\section{Introduction} \label{intro}

Since almost two decades, N-Body (NB) simulations are the best tool
we have to study complex systems such as galaxies and clusters. A
number of different recipes have been proposed to model the gaseous
component of such systems. The best known is probably the Smoothed
Particle Hydrodynamics (SPH) \citep{Lucy77, Benz90}. In brief, it
describes the gaseous fluid with a finite number of particles, which
are followed in their Lagrangian evolution, and evaluates the values
of the physical quantities in each point of the real space by
smoothing each particle through a \textit{kernel} function, peaked
in its position. Together with the equations of hydrodynamics it
allows us to study mechanical heating and shocks; coupled to gravity
described by the well known Tree code (T) by \citet{BarnesHut86},
the NB-TSPH method has proved to be very efficient.

Nevertheless, it suffers some limitations. In particular, it seems
not to be accurate enough to follow in detail the extreme complexity
of the structure and physical processes occurring in a galaxy, which
include radiative cooling, star formation, feedback from
Supernovae, and chemical evolution \citep{Katz92, Mihos94,
ChiosiCarraro02, Merlin06}. These processes, apart from requiring
sufficient computational power in order to achieve the necessary
resolution, also need some kind of different, \textit{ad hoc}
descriptions for the various components. The subject is still at
infancy, which in part explains the many uncertainties currently
affecting the theoretical models. Star formation is usually
expressed by the empiric law of \citet{Schmidt59}, in which the rate
is proportional to the gas density, often coupled with some
conditions reminiscent of the Jeans stability \citep[see
e.g.][]{Gerritsen97, Carraro98, Buonomo00}. If the local gas density
obeys the above conditions, gas particles are supposed to be turned
into star particles. In the so called \textit{sticky-particles}
algorithm \citep{Levinson81, Noguchi86} star formation is considered
to be enhanced by cloud-cloud collisions.

Different recipes have been tested to introduce the complicated
physics of star formation into numerical codes. The most physically
grounded, although expensive in computational load, simply consists
in creating new star particles of different mass according to a
chosen initial mass function (IMF). Obviously this causes a huge
growth of the total number of particles, which is well beyond
present-day technical capability. The situation is more viable if,
assumed a certain mass resolution (initial total baryonic mass in
form of gas divided by the number of mass particles used in the
simulation), a single star particle  is conceived as an assembly of
real stars whose mass distribution follows a given IMF. In this
context, the total number of baryonic particles does not diverge.
Therefore a single gas particle can be turned into a star particle
with no serious consequences for the NB-TSPH ecology. What happens
inside a star particle is relevant only for the energy and chemical
yields but not for the NB-TSPH description as such. This scheme has
been followed in a number of studies. For instance,
\citet{Jungwiert01} considers the gradual formation of stars inside
gas particles, which eventually become star particles when the
hidden star mass exceeds a certain threshold value. Along this line
of thought is  the probabilistic description of star formation by
\citet{Buonomo00, Lia02}, in which gas particles are turned into
star particles according to the Schmidt law however interpreted in
statistical sense, so that at any time there is a certain
probability that a gas particle becomes a star particle. This way of
looking at the star formation process is particularly advantageous
in numerical simulations as it keeps constant the total number of
particles, thus saving lots of computational time.

Owing to the current mass resolution of numerical simulations, the
star particles actually have the mass size of a star cluster, in
which the real stars are born in a short burst of intense local
activity with homogeneous age and chemical composition. Therefore
each star particle can be approximated to a single stellar
population (SSP), in which the stars distribute in mass according to
some IMF.

The influence of the IMF describing the stellar content of each
star particle becomes soon evident when the energy and chemical
feedback  by supernova (SN) explosions and/or stellar winds to the
surrounding medium are taken into account. Several unsettled
questions come soon into play. Among others, it is still not clear
whether the SN energy feedback should be only treated as "thermal"
or if some kinetic effects have to  be considered \citep[see
e.g.][]{Marri03}. How to share chemical elements ejected by a
star particle with the surrounding medium is another point of
uncertainty. Also in this case the statistical description by
\citet{Lia02} turns out to be particularly convenient.

In current NB-TSPH models, the gas component is considered as an
unique fluid regardless of the observational fact that in the gas
content of real galaxies several different phases are present: hot
and cold, tenuous and dense, ionised, neutral and molecular gas.
Furthermore, in the same models star formation and its efficiency
are  related to the total gas content with no attention to the
observational fact that stars are formed only in dense molecular
clouds. In this paper, we present a multi-phase description of the
ISM (hot gas and cold clouds with molecular cores inside) in which
star formation can occur only in the cold component thus bringing
the simulations closer to reality. The new prescription for the ISM
is implemented in \textsc{GalDyn}, the Padua NB-TSPH code originally
developed by \citet{Carraro98} and later improved by \citet{Lia02}
and \citet{Merlin06}.

The layout of the paper is as follows. In Sect. \ref{code} we
present in some details the physical problem and  briefly summarize
our prescription for the ISM and how it is implemented in the
numerical code. In Sect. \ref{test} we test the new method
simulating the formation of an early-type galaxy and present the
results. In Sect. \ref{disc} we discuss the results,  compare them
to some recent observational data on  galaxy formation and
evolution, and draw some general conclusions.

\section{Multi-phase ISM} \label{code}

\subsection{The physical context}
Since the early sixties,  the ISM in the Milky Way is known not to
be uniform on small scales, but to contain strong local density
enhancements, so that very dense and cold regions, in form of atomic
and molecular "clouds", appear to be embedded in a diffuse hot,
ionised medium \citep{Field69}. Some years later, with the
increasing understanding of the importance of SN explosions, another
component was introduced, i.e. the hot and very tenuous coronal gas
expelled by exploding stars \citep{McKee77}.

Therefore, it is nowadays widely accepted that the ISM of the Milky
Way is made of several components with different temperatures and
densities. At least four, almost decoupled, gaseous phases are
present: (i) the hot, highly or fully ionised, tenuous medium, with
temperatures of some tens of thousands of Kelvin degrees and number
densities of about 0.1 cm$^{-3}$; (ii) the cold, dense, clumpy
molecular medium, with temperatures of only a few Kelvin degrees and
number densities from thousands up to millions of atoms per cm$^3$;
(iii) a cloud-embedding, warm, neutral phase, with temperatures from
a few thousands down to a few hundreds Kelvin degrees, and number
density of a few atoms per cm$^3$; (iv) and an extremely hot and
rarefied coronal gas expelled by the SN explosions, with
temperatures of millions of Kelvin degrees and very low number
densities (some 0.01 atoms per cm$^3$). Of course, this scheme is
still a simple one, because the ISM contains other local features
(HII regions, for instance, which are regions of ionised and hot but
very dense material).

In the Milky Way, these different phases fill the interstellar space
in a way that is inversely proportional to their density. Most of
the space is filled by the coronal and hot gas, whereas only a minor
fraction of it is filled by the cold clouds. The cold clouds, on the
other hand, contain the largest fractional mass of the ISM. Owing to
such a strong density contrast, the cold clouds are not strongly
affected by drag interactions with the tenuous phase \citep[see][, 
and  below]{Levinson81}.

Given this situation, it seems clear that the SPH algorithm, which
is fully adequate to describe a single-phase fluid even in
situations of shock, cannot be considered the best tool to study
such a complex mixture of different phases in mechanical and
thermodynamical equilibrium. A well known problem of the SPH method,
for instance, is the so called \textit{overcooling}. Considering the
whole mass of gas as a single fluid, a particle can have
neighbouring particles with much higher densities. These should
actually be considered as part of another decoupled phase. In the
SPH formalism they actually end up to dominate the local 
density of the particle under consideration, thus
leading to overestimate the density and consequently to a stronger,
un-physical radiative cooling \citep[see][]{Marri03}.

Furthermore, the correct equation of state relating
pressure, density, temperature, and chemical composition of the gas
should properly include the molecular component. In most numerical
NB-TSPH codes including our own, the equation of state is that for
an ideal mono-atomic  gas, which  obviously does not adequately
describe the molecular component. Finally, self-gravitating
molecular clouds are known to obey particular empirical laws
\citep[e.g.][]{Larson81}. Likely, they should be taken into account
in the grand-design of star formation.

Over the years, many attempts have been made to solve this problem
and to find a physically sounded recipe for star formation in
molecular clouds. \citet{Levinson81} developed the so called
\textit{sticky particles} algorithm, in which the ISM is modelled
only by means of clouds freely moving across the surrounding void
space. More recently, \citet{Marri03} suggested  a simple
modification of the neighbour searching algorithm in SPH: particles
of very different density and temperature are discarded in the
neighbour sampling. Other strategies have been recently proposed by
\citet{Semelin02}, \citet{Harfst06} and \citet{Booth07}, trying to
simulate the evolution of disk galaxies.

A description of the ISM as detailed as possible is required also to
study  galaxy formation in cosmological context. Even if there are
no evidences that the primordial ISM was similar to the present-day
ISM in the Milky Way, nor  the contrary, it is likely that the ISM
always had an important role. In particular, the duration and
intensity of the star formation activity should be strongly
dependent on how  different components of matter interact each
other. Along this line of thought are the attempts made by
\citet{Springel03} to reproduce the complexity of the ISM in
cosmological context.

\subsection{The multi-phase prescription}
Starting from these considerations and the above mentioned studies,
we tried to elaborate a recipe for the various components of the ISM
and their relationships. In brief,  hot gas is slowly turned into
cold clouds and  star formation occurs in these latter. Viceversa
the energy deposit by dying stars heats up the cold gas. The key
points of our recipe are the listed below.

(1) We assume that the SPH method can reasonably describe the hot
phase of the ISM, considering as "hot" the gas particles which are
hotter and less dense than suitable temperature and density
thresholds (i.e. both the hot ionised gas and the coronal gas).

(2) The cold phase (molecular clouds), in which star formation
occurs, is described by a different method, based on the fact that
individual particles can roughly represent single clumps of cold
material. A giant molecular cloud can reach a mass of some
$10^6\, M_\odot$. Considering also the embedding neutral or
slightly ionised warm material, the total mass of a cold particle
can be higher than this value. As discussed below, the mass
resolution of our fiducial model is close to $6 \times 10^6 \, M_\odot$ thus
making our description quite realistic.

(3) Following \citet{Levinson81}, we assume that the cold clumps can
move almost freely throughout  the hot phase. Pressure drag and
friction are certainly present, but we consider them negligible for
our purposes\footnote{See e.g. \citet{Shu72} for an estimate  of ram
pressure due to the ambient gas.}.

(4) As suggested by \citet{Semelin02}, a hot gas particle does not
instantaneously turn into a cold cloud particle when its temperature
falls below the threshold value because letting this to occur would
originate too sharp a transition. Therefore, we prefer to use a
smooth procedure named \textit{gathering method}. Each gas particle
is thought to be made of gas at different temperatures distributed
according to a gaussian law, truncated at $2\sigma$ from the central
peak determined by the SPH value. To some extent, this mirrors the
situation of the SPH technique, in which each particle has
properties that result from averaging those of nearby particles, so
that only groups of particles can be considered as representative of
the distribution of matter in a given region. The $\sigma_{cold}$ of
this quasi-gaussian distribution is a free parameter. Each particle
is then assigned a certain fraction of cold gas, i.e. the amount of
gas in the quasi-gaussian distribution whose temperature falls below
the threshold value $T_{cold}$. At each time step, a particle tries
to "acquire" some cold gas from its SPH neighbours  until a single
cold particle of mass equal to that of the mother particle is
formed. When this is achieved, a new "cold" particle is created from
the "hot" one. The cold particle is separated from the surrounding
hot medium, provided it belongs to a convergent flow (i.e. its
velocity divergence is negative), and its cooling time is shorter
than its dynamical time, whose evaluation takes Dark Matter around
into account \citep{Buonomo00}.  The temperature $T_{HI}$ of the new
cold particle is determined by the condition that during the
creation process the total thermal energy of the group of particles
remains constant, together with the requirement that neighbours have
no longer cold gas (their temperature distributions accordingly
vary). If the temperature of a gaseous hot particle is limited by
the Jeans threshold (see Sect. \ref{galdyn}, point 3), its
$\sigma_{cold}$ is changed so that it can provide some small
fraction of cold gas anyway. The limitation imposed by the Jeans
threshold is indeed a numerical artifact due to resolution limits,
and there are no real physical reasons to exclude the presence of a
small fraction of cold gas within the total gaseous mass of the
particle. The $\sigma_{cold}$ is adjusted in such a way that the
lower limit of the truncated gaussian distribution
corresponds to zero temperature.

\begin{figure}[htbp]
\centering
\includegraphics[width=7cm,height=7cm]{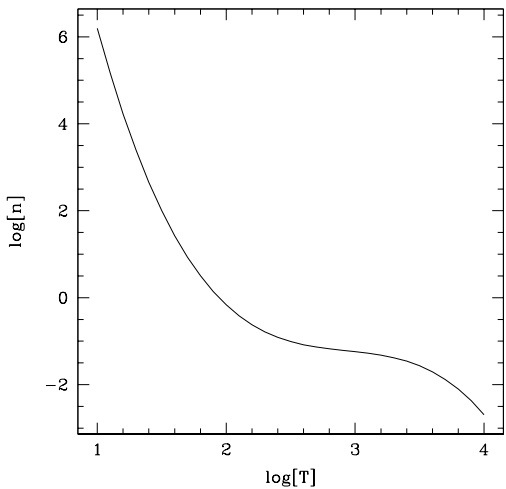}
\caption{Number density versus temperature of atomic Hydrogen in pressure equilibrium.}
\label{rhocloud}
\end{figure}

Finally, we assign the new cold particle the initial density
obtained by interpolating with a cubic polynomial fit the relation
between temperature and number density given by \citet{Scheffler87}
for cold clouds in pressure equilibrium with the surrounding medium:

\begin{equation}
log[n_{HI}] = -0.943x^3 + 8.3x^2 - 24.66x + 23.5,
\end{equation}

\noindent where $x = log[T_{HI}]$. The relation in shown in
Fig. \ref{rhocloud}.

(5) In addition to gravity, the only interactions that cold
particles can feel are the mutual inelastic collisions
\citep{Levinson81}.

Collisions between clouds are known to trigger star formation
\citep{Loren76, Hasegawa94}, due to the creation of a hot shock
front which rapidly cools down causing fragmentation in cold,
extremely dense cores of molecular gas. Numerical simulations of the
impact between two clouds \citep{Marinho00, Marinho01} showed how
the results of such impacts depend on many parameters,  often acting
well below our resolution scale. Anyway, based on those simulations
as well as on the cited observational evidences, we can envision the
following scenario. First of all, two clouds are considered to be
colliding if their mutual distance is shorter then the sum of the
radii of each particle. A rigourous description of
cloud-cloud collisions would require the use of the impact parameter
instead of the distance. However, in the framework of the
statistical approach used for star formation and gas restitution
(see below) using the distance is adequate to our purposes. The
radius of a particle is derived from its mass to total density ratio
and the assumption of spherical symmetry. The total density and mass
include both the neutral and the molecular components. Second, the
densities of the two colliding particles must be comparable. If this
is not the case, the denser cloud will likely cross the less dense
one without feeling sizeable effects. Therefore, we assume that to
have a collision the mean densities of the two colliding clouds
should not differ by more than a factor of ten. Finally, depending
on the mutual distance two kinds of collisions are possible:

\begin{itemize}
\item The \textit{head-on} collision, which
happens if the distance between the clouds becomes shorter then
their mean radius. In this case, the kinetic energies of the
colliding particles are decreased to  $20 \% $ of their original
values, the directions of the colliding components of the velocity
vectors are reversed, and the molecular density of both particles is
increased by a factor of 100.
\item The \textit{off-centre} collision, which
happens if the distance between the clouds gets shorter than the sum
of the two radii but larger than their mean radius. then their mean
radius In this case, the molecular densities are increased
by a factor of 10 and the kinetic energies are decreased to $80 \%$
of their initial value.

\item No interaction is supposed to occur in all other circumstances.
\end{itemize}

\noindent Clouds are allowed to experience collisions until they
reach the molecular density of $10^{-18} \mbox{g/cm}^3$. All
assumptions are consistent with the results of the \citet{Marinho00}
simulations.

(6) The internal evolution of the clumps goes as follows. Each "cold
particle" represents a self-gravitating portion of the ISM in which
two gas phases at different densities and temperatures coexist, i.e.
the molecular core of the cloud and the surrounding  warmer
material. We cannot determine the temperature and density of the two
components, because physical processes at these scales are below the
resolution power of our simulations. It is known that in cold
collapsing clouds the "fragmentation" process can break them into
pieces when the density exceed the value set by the Jeans stability
condition. As already anticipated, we cannot use the perfect gas
equation of state for the molecular part of the clouds, because
different laws are expect to hold \citep[i.e.][]{Larson81} and
complicated processes such as turbulence and magnetic interactions
may play very important roles in the molecular core. To partly get
rid of this difficulty, starting from the study of
\citet{Springel03} in cosmological context, we describe the cold
cloud using mean values that refer to the "molecular part" and the
"neutral" embedding warmer phase, and  assume that in the cold
particle, radiative cooling causes the increase of $H_2$ phase at
the expenses of the $HI$ phase (the process is occurring at constant
temperature for the two components) according to:

\begin{equation}
\frac{\delta\rho_{H_2}}{\delta t}=-\frac{\delta\rho_{HI}}{\delta t}=
\frac{\Lambda_{HI}(\rho_{HI},u_{HI})}{u_{HI}-u_{H_2}} \rho_{HI}.
\end{equation}

In practice, when a gas particle detaches itself from the
surrounding hot ambient medium and becomes a "cloud", it is thought
to be made of molecular and neutral gas in some proportions (in
our simulations, we assume that no $H_2$ is present when the cloud
is formed and it increases afterwards). The radiative losses "cool"
the neutral part and the "cooled" material is subtracted from it and
accreted onto the molecular part, thus changing the densities of the
two phases (at constant occupational volume).

(7) Only cold particles are allowed to form stars. The transition
from cold gas particles to star particles is based, as
explained in \citet{Lia02}, on a probabilistic interpretation of the
\citet{Schmidt59} law. In practice, we consider that on a time scale
of about 3 Myr (the lifetime of a typical high mass star) a fraction
of the molecular gas, proportional to its density, is turned into
stars according to

\begin{equation}
\frac{d\rho_*}{dt}=c_*\frac{d\rho_{H_2}}{t_{dyn}}.
\end{equation}

Instead of creating more and more new particles of different mass,
which would make the computational cost too high, we prefer to
interpret this formula as the probability that a cloud particle is
completely turned into a star particle. The  probability is

\begin{equation}
P(SF)=\left[\frac{\rho_{H_2}}
{\left(\rho_{H_2}+\rho_{HI}\right)}\right]\left[1-exp\left(\frac{-c_*\Delta
t}{t_{dyn}}\right)\right]\, ,
\end{equation}

\noindent where the first term takes into account that we consider
only the molecular density when turning \textit{the whole particle}
into stars. \citet{Buonomo00} showed that, provided the total number
of particles is sufficiently high (say more than $ \simeq 2000$),
the right global behaviour is achieved (the statistical description
equals the straight one). Following \citet{Lada03}, we adopt
$c_*=0.05$. As already mentioned, close encounters between clouds
enhance star formation, due to the growth of molecular densities
after a collision.

(8) Along with radiative cooling and energy dissipation by inelastic
collisions, the thermal evolution of clouds is governed by the
energy input by SN explosions. The SN energy feedback is perhaps
the most difficult issue to address, because of its complex and
poorly known physics. Furthermore, all physical processes concerning
SN explosions and evolution of the SN remnants (SNR)  are well below
both our time and space resolution scales (namely, $\sim$ 1 Myr and
$\sim$ 1 kpc), so that we are forced to represent them through
"global" changes on large ISM regions. Here we  adopt a simple
description and try to minimize the number of free parameters.

feedback energy from SN explosions is distributed among particles
as follows. The "death" of star particles  and associated
SN explosions are described with the statistical method proposed by
\citet{Lia02}, see also \citet{Merlin06}. In brief, a star
particle (a SSP of a certain age) is sorted out through a Monte-Carlo
method and its "death probability" is measured by the ratio between
the amount of mass lost by stellar winds and SN explosions and
total SSP mass. The probability goes from zero to one when the ratio
goes from "zero to one" (from zero to a maximum value that
ultimately depends for a given IMF on the mass interval within the
SSP in which SN explosions are supposed to occur). In this way, we
avoid instantaneous recycling and, at increasing number of star
particles, we get closer and closer to a realistic description.
When a star particle dies, its
mass is completely turned back to the hot phase, and the energy
feedback due to SN explosions within the SSP is spread over the
particle itself and its neighbours, in the following way. First
of all we compute the total gaseous mass $m_{tot}$ inside a sphere
of radius equal to the smoothing length of the newly born gas
particle\footnote{Using the particle smoothing length instead of a
fixed distance causes feedback to be differently efficient in
different environments, and avoids the introduction of parameters.}.
$m_{tot}$ is thus given by the mass of the particle itself plus the
mass of all cold clouds inside the sphere of  radius equal to the
smoothing length. The corresponding release of specific energy $u$
[$erg/g$] is

\begin{equation}
u = \frac{N_{SN}E_{th}}{m_{tot}},
\end{equation}

\noindent where $N_{SN}$ is the number of SN explosions within the
dying particle, calculated according to the \citet{Greggio83} rates,
and $E_{th}$ is the thermal energy produced by each explosion (see
below). Each cloud is then given an amount of thermal energy
corresponding to its mass, decreased accordingly to the smoothing
kernel of the newly formed gas particle, and the exceeding energy is
given to the new gas particle and added to its original thermal
energy. Furthermore, thermal energy is spread among the hot
neighbouring particles according to the normal SPH algorithm.
Following \citet{Thornton98}, we assume that the amount of thermal
energy spread over the ISM by each SN explosion (which yields a
total energy of $10^{51}$ ergs per event) depends on the ambient gas
conditions. Thus, we compute the mean values of density and
metallicity $n_0$ and $Z/Z_{\odot}$ of the neighbouring particles,
weighing each contribution on the smoothing kernel, and use the
following relation to obtain the total amount of thermal energy to
be spread into the surrounding medium \citep[see][ for
details]{Thornton98}:

\begin{displaymath}
E_{th} = 1.83 \times 10^{49} n_0^{-0.23} (Z/Z_{\odot})^{-0.24}
\end{displaymath}
\begin{equation}
       + 1.23 \times 10^{49} n_0^{-0.24} (Z/Z_{\odot})^{-0.08} ergs,
\end{equation}

\noindent if $log[Z/Z_{\odot}]>-2$, and

\begin{displaymath}
E_{th} = 5.53 \times 10^{49} n_0^{-0.23}
\end{displaymath}
\begin{equation}
       + 1.78 \times 10^{49} n_0^{-0.24} ergs
\end{equation}

\noindent otherwise. Finally, each cloud is given an amount of
radial kinetic energy, still weighed on the kernel and computed in
the same way as for the thermal energy, using the empirical relation
$E_{kin} = 1.69 \times 10^{50}$ ergs of \citet{Thornton98}.

With this mechanism, clouds are allowed to "evaporate" back to the
hot phase as soon as the temperature of their neutral part  gets
higher than a threshold value, that is naturally taken equal to the
cloud formation temperature. In addition to this, we try to mimic
the first phase of adiabatic expansion of the SNR and passage of the
compressional front \citep[see][ for details]{McKee77} by
momentarily inhibiting radiative cooling in clouds which experience
SN heating. By doing this we ensure that sudden cooling of the
particles cannot take place simply because soon after the explosion
they happen to be immersed in a denser environment.

(9) Finally, the chemical evolution  of the ISM is followed as in
\citet{Lia02}. It is worth noting, however, that \textit{cold
particles are supposed  not to experience chemical enrichment},
which is limited to the hot medium. Only when a cloud has been
evaporated by SN energy feedback it can enrich its gas with metals.

\begin{figure}[htbp]
\centering
\includegraphics[width=9cm,height=8cm]{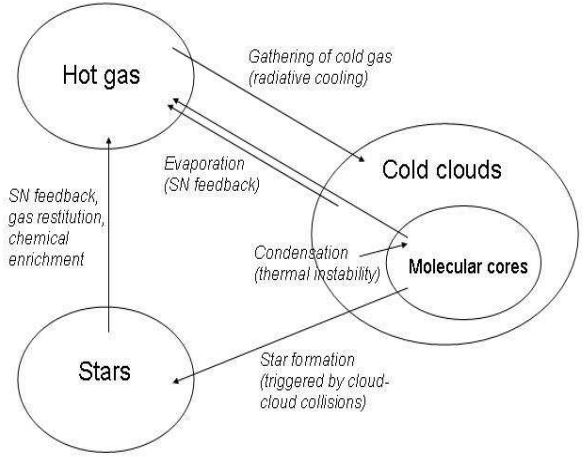}
\caption{The different interactions among the three baryonic
components of a galaxy with the multi-phase description of the ISM.}
\label{inter}
\end{figure}

Fig. \ref{inter} shows a schematic summary of the different
interactions among the baryonic components of a model galaxy with
the new scheme.

\section{Testing the new recipe} \label{test}

\subsection{GalDyn, The Padova NB-TSPH code} \label{galdyn}

\textsc{GalDyn} is an evolutionary code for galactic systems, developed
in Padova over the years. It is a fully
lagrangian NB-TSPH code, based on the well known Tree-code by
\citet{BarnesHut86} and SPH algorithm by \citet{Lucy77}. For a
detailed description of the basic version of the code see
\citet{Carraro98, Buonomo00, Lia02, Merlin06}. For the purposes of
this study, the code has been improved in several ways with respect
to the previous versions.

(1) First of all, an adaptive multiple time-step integration scheme
has been adopted. For each particle an individual time-step taken as
the minimum between the acceleration time-scale, the velocity
time-scale, and, for gas particles, the Courant time-step
\citep[see][ for details]{Merlin06} is calculated using its current
physical values. Within each time-step for the whole system, only
those particles with a sufficiently small individual time step are
considered to be "active" and are moved on, using a Drift-Kick-Drift
Leapfrog integrator. This is similar to the implementation used  by
\citet{Springel01} in \textsc{Gadget 1}. However, it is worth noting
that the Drift-Kick-Drift algorithm  may  suffer non negligible
problems for what concerns the conservation of the motion integrals
 \citep[see][]{Springel05}. Note that particles are not allowed to
have a time-step lower than a minimum value (fixed equal to
$10^{-5}$ of the Hubble time), and whenever required their smoothing
length is changed to establish the right value.

(2) The adopted smoothing kernel for SPH particles, originally
developed by \citet{Monaghan85}, has been revised following
\citet{Thacker00} and references therein. In order to avoid
artificial clumping in high density regions, the first derivative of
the kernel function is set  constant and equal to $1/(\pi h^3)$ for
small radii.

(3) Radiative cooling is not allowed to be active on a particle when
its temperature falls below the threshold value given by

\begin{equation}
T_{min} = C \left[
\left(\frac{4 m_p}{\rho_g}\right)^{2/3}-\left(\frac{\pi
\epsilon_g^3}{3}\right)^{2/3} \right] \frac{\mu m_H G}{k}\rho_g,
\end{equation}

\begin{figure}[htbp]
\centering
\includegraphics[width=7cm,height=7cm]{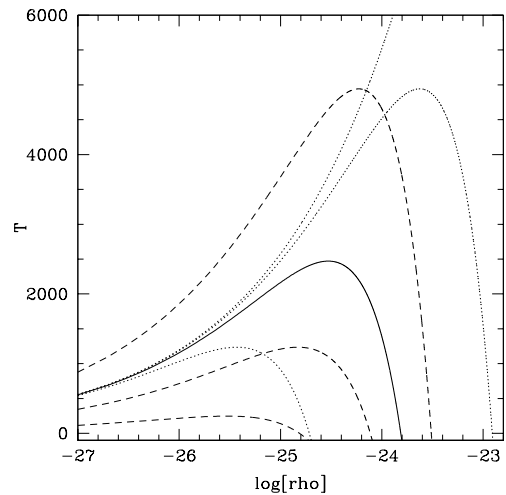}
\caption{Lower threshold temperature for SPH particles of mass $m_p$ and
softening length $\epsilon$, as a function of their density. Solid line:
$m_p$ = $m_{0,p}$ = $6 \times 10^6$ M$_{\odot}$, $\epsilon$ = $\epsilon_0$
= 1 kpc. Dotted lines: from top to bottom,
$m_p$ = $m_{0,p}$ and $\epsilon$ = 0.1, 0.5, 2 times $\epsilon_0$.
Dashed lines: from top to bottom, $\epsilon$ = $\epsilon_0$ and $m_p$ =
2, 0.5 and 0.1 times $m_{0,p}$.} \label{jeans}
\end{figure}

\noindent \citep[see][]{Bonnor56} where $m_p$ is the mass of a
baryonic particle, $k$ is the Boltzmann constant, $\mu$ and
$\epsilon$ are the mean molecular weight and the gas softening
parameter, $m_H$ is the proton mass, and the C is a constant (in
\textit{cgs} units, C=0.89553). Below this temperature the particle
would become Jeans unstable (obviously, adiabatic cooling is not
limited by this condition). This somehow imposes an unavoidable
limit to the mass resolution of the particles because, if the
threshold value is too high, cooling and cloud formation cannot be
correctly modelled. In Fig. \ref{jeans} the threshold temperature in
Kelvin degrees is plotted versus the logarithm of density (in
g/cm$^3$), for several values of the particle masses and softening
lengths. The solid lines refers to the values adopted in the present
simulations, see Sect. \ref{mod}.

(4) In order to correctly evaluate the impact of the inverse Compton
cooling at high redshifts, we roughly estimate the ionization
fraction and thus the number density of free electrons as a function
of temperature. The number density is

\begin{displaymath}
n_e = n_{HI} \mbox{  if  } T \geq 10^4 \mbox{  K,}
\end{displaymath}
\begin{equation}
n_e = n_{HI} \times exp(-(10^4-T)) \mbox{  otherwise,}
\end{equation}

\noindent where the value of $10^4$ K roughly corresponds to the
ionization temperature  of Hydrogen in astrophysical environments.

(5) We describe the thermal conduction among SPH particles as
originally developed by \citet{Monaghan91}. Thermal conduction is
expected to be important in situations of shocks, and in particular
after a SN event. \citet{McKee77} pointed out that thermal
conduction could evaporate cold clouds and warm up the  ISM inside a
SNR  (thus also changing the density inside the remnant). Although
in our scheme clouds are evaporated "automatically" due to SN
events, we expect thermal conduction to be important in the
subsequent phases of evolution, when a large number of hot particles
are present in a high density region.

(6) To take into account UV flux and stellar winds from massive
stars without too much computational effort, we make use of  the
results of \citet{Buonomo00} who found that the contribution to the
energy feedback by these two sources of energy is comparable to
that of SN. Since stellar winds and UV flux come the same stars that
eventually will explode as Type II SN, we crudely lump together the
three contributions and simply increase by a factor of 3 the energy
injection by SN explosions.

\subsection{Formation  of a galaxy in cosmological context} \label{mod}

We have tested the new prescription for star formation
implemented in  \textsc{GalDyn} by simulating the formation of a
galaxy from quasi-cosmological initial conditions.

The cosmological scenario we adopt is consistent with the W-Map3
results \citep{Spergel06}, namely $h_0 = 0.73$, $\Omega_m = 0.238$,
$k = 0$, $\sigma_8 = 0.74$, $n = 0.951$, and initial baryon fraction
$ \sim 17.6\%$.

The initial conditions are obtained using the \textsc{Grafic2}
package by \citet{Bertschinger01} to generate an initial
perturbed cubic grid of 64$^3$=262,144 particles. In this grid we
search a region containing an overdensity peak and single out
another much smaller cubic volume around it, so that the peak is
located near its geometrical centre. This small cubic region is
initially made of 6$^3$=216 particles (corresponding to a cubic
volume of $\sim 20^3$ physical kpc$^3$) and is then re-simulated
with a finer resolution, which is imposed to give at the end a
number of particles compatible with our computing capability.
Finally, in the re-simulated small cubic volume we isolate the
sphere whose radius is half of the  cube dimension and whose centre
coincides with that of the cube (i.e., with the original position of
the density peak in the large cubic grid). This final sphere
contains $\sim$ 14,000 particles (approximately half of dark matter
and half of baryons). The sphere is inserted into the conformal
Hubble flow with void boundary conditions. These latter may not be
fully adequate to describe the real situation because the isolated
volume is in reality immersed in a background exerting some
dynamical effects on the matter contained inside the sphere under
consideration. To cope with this, we add a small initial solid-body
rotation to the sphere, and follow its expansion, detaching from the
Hubble expansion, turn-around, and collapse \citep[see e.g.][ for
details on the initial conditions]{Merlin06}. The sphere can be
considered as a good approximation of an isolated initial
proto-galaxy. Despite the small solid body rotation, the void
boundary conditions may cause some problems, which anyway are not
crucial for our purposes \citep[see][ for more details]{Merlin06}.

As already said, the way the cubic volume single out  from original
grid has been re-simulated is chosen in such a way that the total
number of particles in the so-called proto-galaxy  is compatible
with our limited computing facilities. We are well aware
that the rather small total number of particles may constitute a
major drawback of the results we are going to present. The main
motivation for this choice is that we want to test the new scheme of
star formation with an affordable computing cost and time.
However, several studies \citep[see e.g.][]{Kawata99,
Buonomo00, Kawata01a,Kawata01b,Kawata03,Semelin02} have already
shown that a total number of particles of the order of $10^4$ is
enough to achieve a sufficiently accurate global description of the
physical processes. Therefore, we are confident that the main body
of the physical conclusions we are presenting are not too badly
influenced by resolution limits. They have to be viewed as
indications of the right trail to follow in the future.

We also consider the action of the re-ionizing background radiation
(e.g. the UV field of the PopIII massive stars). To deal with effect
in a very simple way, we crudely  add a constant heating of
$10^{-24}$ ergs/s/atom from redshift z=20 to z=6.

As the simulation starts well before re-ionization and the  initial
overdensity of the proto-galactic halo is small, Hydrogen can be
considered as almost completely neutral at the initial stage.
Anyway, the initial temperature of the gas is set equal to the
maximum between a "cosmological" value, obtained as explained in
\citet{Merlin06}, and the minimum temperature allowed by the Jeans
limit discussed above. Indeed, the cosmological temperature
of baryonic matter at $z \sim 60$ would be below 100 K, but the
Jeans resolution in the present models imposes a threshold value of
$\sim 2000$ K. As the Jeans threshold  is  well below the value at
which ionization is expected to yield sizeable effects, we can
safely use this value instead of the real cosmological limit. In
any case, at the very initial stage we must consider  the gaseous
component as made only by "hot" particles, because as already explained,
cold clouds are allowed to form only when belonging to collapsing
flows.

\citet{Ballesteros06} estimated the following  densities and
temperatures for the two main components of a cold cloud:  $1<n<100$
cm$^{-3}$ and $100<T<5000$ for the atomic gas; $100<n<1000$
cm$^{-3}$ (up to $10^6$ in cores) and $10<T<100$ K for the molecular
gas. Therefore,  we adopt $T_{cold}=5000$ K and
$\rho_{cold}=10^{-27}$ g/cm$^3$  as threshold values. Such a low
value for the density threshold is motivated by the fact that this
is the global SPH density of a particle, while in order to create
cold clouds we are only interested in the densest part of it. We
then choose $\sigma_{cold} = 1000$ K. Finally, to describe
the properties of of our star particles (energy feedback and
chemical enrichment) we adopt the \citet{Salpeter55} IMF.

Our model proto-galaxy has an initial radius of $19$  kpc, and a
total (Dark and Baryonic) mass of $2.5 \times 10^{11} M_{\odot}$,
with an overdensity of $\simeq 0.09$ at the initial redshift $z=60$.
The number of particles for  Baryonic  and Dark Matter is $\simeq
7000$, the same for each component. For baryons the  mass of a
particle is  $\simeq 6 \times 10^6 M_{\odot}$, and the softening
length is 1 kpc; the corresponding values for the Dark Matter
component are $\simeq 3 \times 10^7 M_{\odot}$ and 2 kpc, because of
the much higher total mass. All the relevant quantities in use are
summarized in Table \ref{tab1}.

\begin{table}
\caption{Initial parameters for the model proto-galaxy} \centering
\begin{tabular}{|l|l|}
\hline
Initial redshift                            & 60            \\
\hline
Initial mean overdensity                    & 0.09           \\
\hline
Initial radius [kpc]                        & 19           \\
\hline
Total mass [$M_{\odot}$]                    & $2.5 \times 10^{11}$  \\
\hline
Number of baryonic particles                & 7000        \\
\hline
Mass of a baryonic particle [$M_{\odot}$]   & $6 \times 10^{6}$     \\
\hline
Softening length for a baryonic particle [kpc]      &  1 \\
\hline
Number of Dark particles                    & 7000      \\
\hline
Mass of a Dark particle [$M_{\odot}$]       & $3 \times 10^{7}$     \\
\hline
Softening length for a Dark particle [kpc]      &  2 \\
\hline
\end{tabular}
\label{tab1}
\end{table}

\begin{figure}
\centering
\includegraphics[width=7cm,height=7cm]{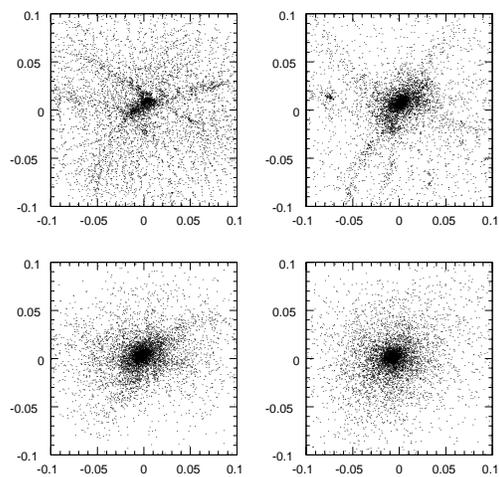}
\caption{From top to bottom, from left to right: position (physical Mpc) of Dark
Matter particles projected on the XY plane at redshift $z=5.9, 3.4,
1.7, 1.0$.} \label{pos_cdm}
\end{figure}

\begin{figure}
\centering
\includegraphics[width=7cm,height=7cm]{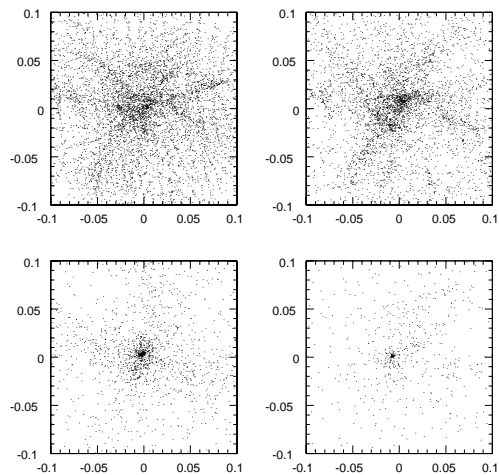}
 \caption{Same as above, but for hot gas particles.} \label{pos_hot}
\end{figure}

\begin{figure}
\centering
\includegraphics[width=7cm,height=7cm]{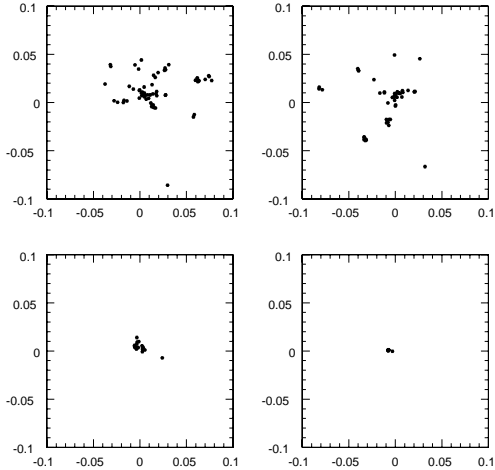}
\caption{Same as above, but for cold gas particles.}
\label{pos_cold}
\end{figure}

\begin{figure}
\centering
\includegraphics[width=7cm,height=7cm]{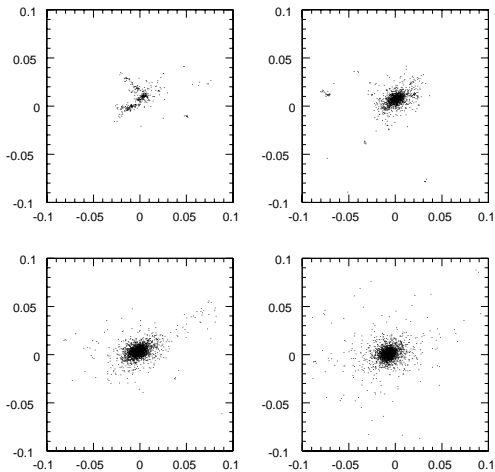}
\caption{Same as above, but for star particles.} \label{pos_star}
\end{figure}

\begin{figure}
\centering
\includegraphics[width=7cm,height=7cm]{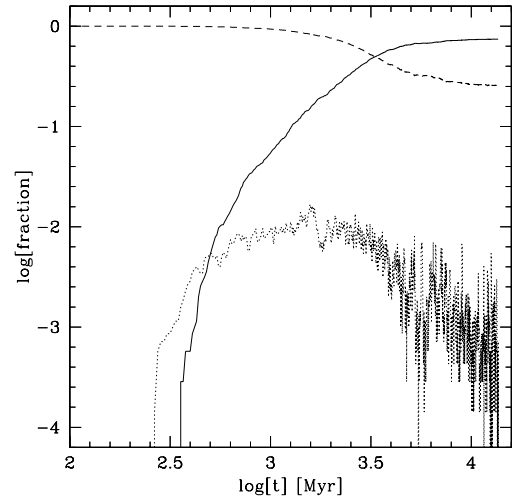}
\caption{Fractional mass in the three baryonic components versus
time. Dashed: hot gas; dotted: cold clouds; solid: stars.}
\label{frac}
\end{figure}

\subsection{Formation and evolution of a galaxy: results and comments}

First of all, we  note that owing to the present procedure to
isolate the proto-galaxy from the cosmological context, the void
boundary approximation may influence the evolutionary time scales
and the detailed results in turn. In particular, it may (slightly)
change the relationship between rest-frame evolutionary time and
redshift \citep[see][]{Merlin06}. Despite this cautionary remark, we
are rather confident that the basic features and properties of the
model should not depend very much from this drawback of the present
formulation.

\begin{figure}
\centering
\includegraphics[width=7cm,height=7cm]{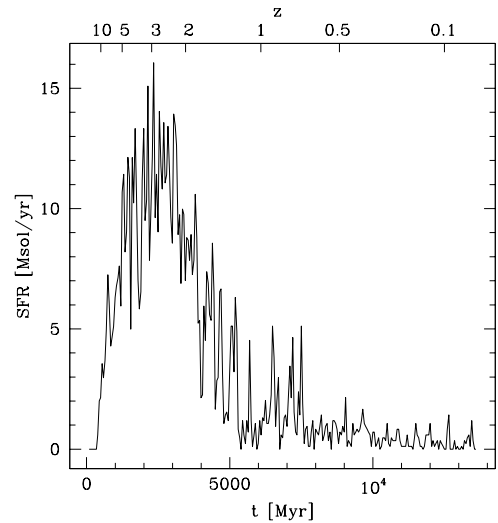}
\caption{Star formation rate in solar masses per year.} \label{sfr}
\end{figure}

\textbf{Galaxy formation}. In Figs.  \ref{pos_cdm} through
\ref{pos_star} we show the spatial evolution of the system, in its
four components: Dark Matter, hot gas, cold clouds and stars. All
boxes are 100 kpc per side. The number of cold clouds (whose
dimensions have been magnified for clarity) is always quite small,
as they soon evolve into stars or are  evaporated back to the hot
phase; this is clearly visible in Fig. \ref{frac} showing the
fractionary mass of the baryonic components as a function of the
cosmic time. This situation, which differs from the one in a disk
galaxy like the Milky Way, where most of the gaseous mass is stored
in the cold phase, can be explained as follows. First, there is a
technical reason: as already noted, at the beginning of the
simulation, in order to apply the SPH formalism and get the right
values for densities and temperatures, we consider the gaseous
component to be made only by  hot particles, even if their
temperature is below the threshold value for cloud formation. This
effect  should become less relevant as the galaxy ages. Second, the
mass resolution is small, so that statically the number of cold
clouds going into stars per time step is also small (and
fluctuating). The third reason is more physical: in our model galaxy
(simulating what happens in an elliptical) a large fraction of gas
is  turned into stars so that little gas is left after a few Gyr of
star formation activity (see below). Most of the cold gas sinks very
early toward the centre of the proto-galaxy, whereby undergoing
frequent collisions is turned into stars at a high rate. Dying stars
soon provide SN feedback to the surrounding medium, thus inhibiting
the formation of a great number of new cold clouds. Conversely, in
situations with low star formation rates (such as in evolved disk
galaxies) the cold gas is expected to distribute almost uniformly
across the disk and likely to co-rotate with the rest of the galaxy;
the number of cloud-cloud collisions is of course lower and lots of
cold gas survive to star formation. See below for some
additional comments on this issue.

The clumps of collapsed matter merge at very early times, and at $z
\simeq 3.5$ the main body of the final object is in place, accreting
new material from the outskirts. In the context of current
terminology in literature, we prefer to classify this situation as
"monolithic" rather than "hierarchical". In any case it would be a
"very early hierarchical" assembly of a galaxy\footnote{We name
\textit{monolithic} a formation process in which all the relevant events
which give a galaxy of any size its final shape and features, e.g. mergers and
dynamical evolution, have taken place in the remote past, say before
z $\simeq 2$, in contrast to a hierarchical view in which bigger
objects are assembled from smaller objects over the whole Hubble
time.}. At the end of the simulation ($z \simeq 0$), the galaxy has
a nearly spherical shape with axial ratios $b/a = 1.02$ and $c/a = 1.12$.
The total stellar mass is $\simeq 3 \times 10^{10} M_{\odot}$ ($75\%$ of the
initial baryonic mass), with an half mass radius $R_{1/2} = 4.6$ kpc. Almost
no gas is left inside the virial radius of the object.

In Fig. \ref{sfr} the rate of star formation (in solar masses per
year) is shown. The first stars begin to form around $z \simeq 12$.
The bulk of stars (80\% of the total) is in place at $z \simeq 1.6$.
The galaxy has already acquired its final features, and from now on
it evolves almost passively. The last, minor burst of star formation
takes place at $z \simeq 0.7$. These result is in good agreement
with the observations for objects with similar total stellar mass
\citep[see e.g. ][]{Thomas05, Clemens06, Cimatti06}. In the nowadays
well-established \textit{downsizing} scenario, smaller objects
assemble their stellar content over longer time scales than massive
galaxies - at least in high density environments. Among others,
\citet{Jimenez05} studied a large sample of $\simeq 10^5$ galaxies
from the SDSS, and found a tight relation between the total stellar
mass of an object and epoch of its assembly (see e.g. their Fig. 4).
A system with $10^{10} < M_* / M_{\odot} < 10^{11}$ is expected to
form most of its stellar content about $\simeq 5 -6$ Gyr ago, whilst
more massive objects are expected to exhaust their activity much
earlier (the situation being more intrigued for low mass objects,
which have different histories depending on their initial
overdensity). Our results are in very good agreement with
this picture, and no particular fine tuning of parameters was required
to obtain them. Indeed, all the parameters are conservative and
taken from previous studies of both observational and theoretical
nature. Perhaps, a more prolonged period of activity could have
been expected for our object, given its mass and its rather low
initial overdensity, but of course we cannot too strictly compare
the results for a single model galaxy with the statistically
averaged trends that are found observationally.

\begin{figure} 
\centering
\includegraphics[width=7cm,height=7cm]{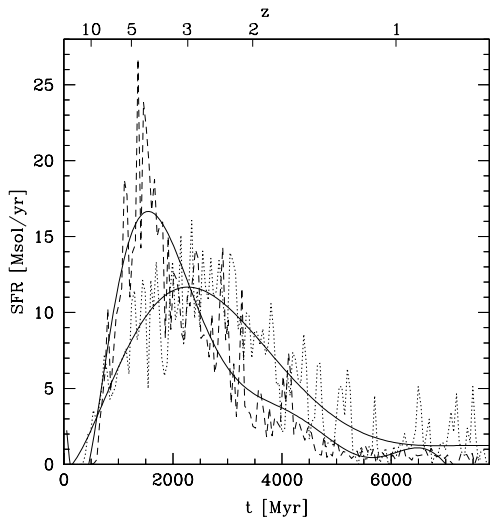}
\caption{Star formation histories. Dotted: the new multi-phase
description of the ISM; dashed: the standard SPH model.
The solid lines are best fitting polynomial functions, see text for details.} \label{sfr_comp}
\end{figure}

In order to achieve some insights on this point, as well as on the
improvements due to the new prescription for star formation, we
compared the present results with those obtained with the classical
SPH treatment of the ISM \citep[see][ for details]{Merlin06}. The
two star formation histories are shown in Fig. \ref{sfr_comp}, where
the solid lines are polynomial best fits of the numerical results,
sketched in order to give an idea of the trends of the star
formation histories in the two models apart from the very
fluctuating profiles, which are due partly to numerical artifacts
and partly to the statistical method we have adopted to deal with
star formation and energy injection by dying stars. The SPH
simulation was stopped after 7 Gyr of evolution due to limited
computational resources. With the new prescription, the formation of
the stellar mass is more gradual and lasts longer; the peak of
activity is lowered from $\simeq 17$ to $\simeq 12$ solar masses per
year, and shifted from $z \simeq 4.3$ to $z \simeq 2.9$. The total
stellar mass assembled in the two models is, however, almost equal.
The different behaviour obtained with the two star formation
prescriptions can be understood first of all by looking at the
different time scales involved in the formation of new stars. In the
classical SPH treatment, all gas particles can in principle be
turned into stars, and therefore a large reservoir of fuel is ready
to be used right from the beginning, giving rise to a strong and
sudden burst of activity. In contrast, with the new scheme, the gas
has first to generate  cold clouds, than these must develop a
molecular core, and finally the core must reach a high density
before a star particle can be born (only the molecular part of the
cold particle is considered in the statistical algorithm adopted to
form stars). Another difference between the two models is
the amount of energy injected into the gaseous component. While in
the classical SPH treatment only the SN feedback was taken into
account, in the new recipe we also considered other sources such as
UV flux from young stars, stellar winds from massive stars, and the
kinetic kick to cold clouds by SN explosions. As widely discussed
below, this has certainly sizeable effects on the star formation
histories of the two models. All this end up in a longer and
shallower period of activity. Of course, being the law of star
formation at work always the same but for the amount of material to
its disposal, the change in the global time variation is not
dramatic, though sizeable.

\begin{figure}  
\centering
\includegraphics[width=7cm,height=7cm]{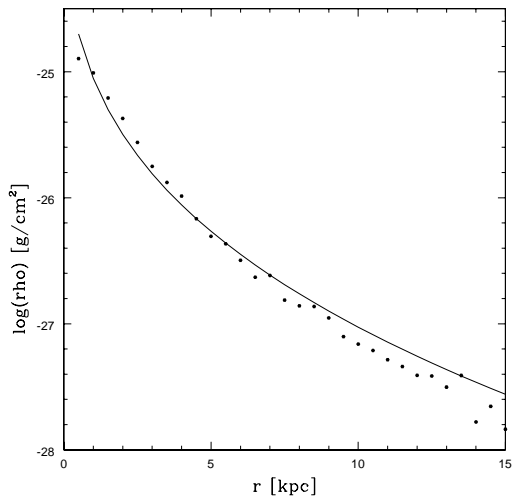}
\caption{Surface density profile of star particles. The solid
line is a fitting Sersic profile ($m$ = 3).} \label{vauc}
\end{figure}

\begin{figure}   
\centering
\includegraphics[width=7cm,height=7cm]{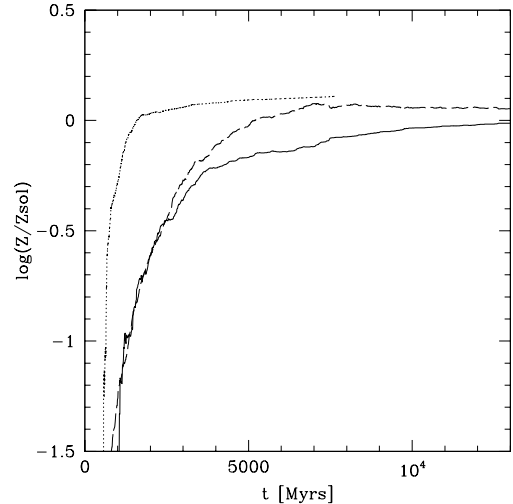}
\caption{Mean chemical composition versus time. Solid: stars;
dashed: hot gas; dotted: stars with the old description of the ISM.}
\label{chemo}
\end{figure}

\textbf{Tails of star formation}. The  model shows a very long tail
of stellar activity in the very central regions which might be taken
as a flaw to be eliminated. However, some recent studies \citep[see
e.g.][]{Bressan06,Clemens06} argue for the presence of a small tail
of ongoing star formation up to low redshifts, at least in clusters,
for a conspicuous fraction of early type galaxies. With the
present model, whether the tail corresponds to that suggested by the
observations cannot be said because of the the mass resolution.
Indeed, if the number of gas particle turned into stars during the
tail activity is very small (of the order of one to two particles
per time step) the mass of newly created stars is always significant
because of the mass resolution. We expect that increasing the mass
resolution (initial total number of particles) should clarify the
issue and lead to quantitative estimates of the star formation
efficiency in the tail activity (if any). On the other hand, a
mechanism able to completely stop star formation in most massive
early type galaxies is still missing. Many hints suggest that the
presence of an AGN could be the solution. Work is in progress to
address the problem.

Finally, we would like to note that the  mass in stars of our model
is almost equal to the limit of $3 \times 10^{10} M_{\odot}$ found
by \citet{Kauffmann03} and \citet{Jimenez05}, between "red and
massive" and "blue and small" objects in the SDSS. In this sense, we
could expect to obtain an object with intermediate features between
the two galactic populations: the star formation history of our
model goes in the right direction.

\begin{figure}  
\centering
\includegraphics[width=7cm,height=7cm]{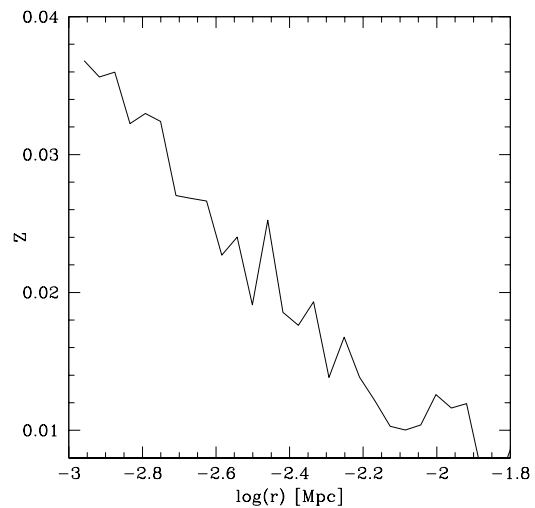}
\caption{Metallicity gradient for star particles.} \label{met}
\end{figure}

\textbf{Surface mass density profile}. In Fig. \ref{vauc} we show
the surface density profile of the final stellar component together
with a \citet{Sersic68} profile with $m = 3$. The profile fits quite
well the Sersic relation, especially if we consider that our profile
is actually a projected mass profile, which should be
photometrically translated into a luminosity profile in order to
obtain a better comparison with the observational profiles.

\textbf{Chemical enrichment}. Fig.\ref{chemo} presents the
metallicities of stars and hot gas. The final value for the star
metallicity, at $z \simeq 0$, is solar within a few percent, in
striking agreement with the results by \citet{Gallazzi05}. By
contrast, in the "old" model the metallicity is already super-solar
at $z \simeq 4$, and at $z \simeq 0.7$ is 1.3 Z$_{\odot}$, which is
expected for objects ten times as massive (dotted line in Fig.
\ref{chemo}). Likely, this is partly due to the same
mechanism that decreased the star formation rate, and partly  to the
procedure adopted to spread metals into the interstellar medium. As
already explained, metals are indeed distributed only among hot gas
particles, leaving cold clouds with their own original metallicity
until they are heated up back to the hot phase by SN feedback, thus
delaying the chemical enrichment of the material out of  which stars
are formed.

\textbf{Metallicity gradients}. Fig. \ref{met} shows the radial
metallicity gradient of the stellar component. It has been long
known that elliptical galaxies are optically redder in their cores
\citep{Vauc61}; this has been interpreted in terms of higher
metallicity  due to prolonged star formation in the central regions.
This is well achieved in our model.

\begin{figure}
\centering
\includegraphics[width=7cm,height=7cm]{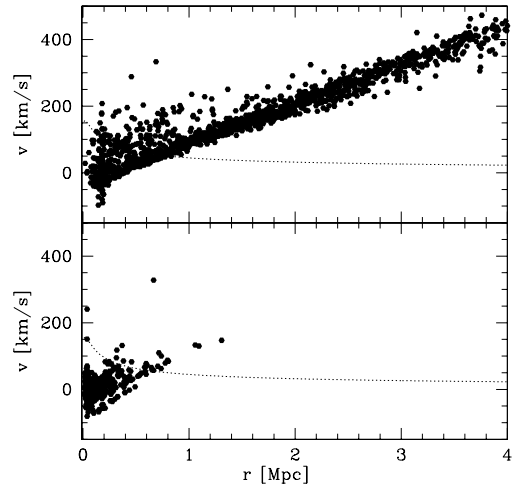}
\caption{Upper panel: mean radial velocity of gas particles versus
radial distance, compared to the escape velocity at the same
distance (dotted line). Lower panel: same as above, for a model with
SN feedback nearly switched off.} \label{gasfuga}
\end{figure}

\textbf{Galactic winds}. Finally, conspicuous galactic winds are
found to occur. The upper panel of Fig. \ref{gasfuga} shows the mean
radial velocity of the gas particles at z $\sim 0$, as a function of
their radial distance, compared to the escape velocity at the same
distance. For the sake of comparison, the lower panel shows the same
plot for an identical model where energy feedback had been
drastically decreased (almost switched off, see below): the number
of gas particles are fewer (one third of total gaseous mass), and
almost all of them are confined within the galaxy potential well. To
avoid possible misunderstanding of the physical meaning of Fig.
\ref{gasfuga} it is worth commenting on the particles that are found
at very large distances from the mother galaxy. This is a result of
the void boundary conditions and the computational scheme in use.
Particles that acquire velocities larger than the escape velocity
and are free to travel without feeling other effects can indeed
reach very large distances (for instance a particle escaping at
about 400 km/s and keeping this velocity constant for about 10 Gyr
travels up to $\sim 4$ Mpc). Second the numerical code keeps track
also of these particles even if they do not longer belong to  the
galaxy. Fig. \ref{gasfuga} simply shows that large amounts of gas
can be considered unbound to the galaxy, because of the energy
feedback, and transferred to the  external environment as a galactic
wind. This has slightly super-solar mean chemical composition and
will affect the chemical composition of the medium in which galaxies
are immersed, the well known issue of the chemical composition of
the intra-cluster medium \citep[see][\, for recent discussions of
this subject]{Chiosi2000,Moretti2003,Portinari2004}. The results for
the model with drastically decreased energy input also deserves some
comments. In this models the energy input is not completely switched
off because otherwise all gas particles around would have been
turned into stars and no gas would have been left over. The energy
decrease is modulated in such a way that the energy input brings gas
particles to the temperature of about 50,000 K, i.e. half of
the hydrogen first ionization potential (13.6 eV per atom) and well
above the temperature at which the astrophysical low-density plasma
starts ionizing and below which it begins to form cold clouds.
Compared to the case with full energy input, in which the
temperature of the gas particles can reach some $10^7 - 10^8$ K,
there is a difference of at least a factor of 100. The kinetic
energy acquired by the gas particles scales with the square root of
the temperature and therefore the cloud velocities are now about a
factor of 10 smaller than in the full energy input cases. Over a
time interval of 10 Gyr these particles can travel about 1 Mpc and
some of them can even acquire sufficient energy to exceed the escape
velocity, consistently with the results shown in Fig. \ref{gasfuga}.

\begin{figure}
\centering
\includegraphics[width=7cm,height=7cm]{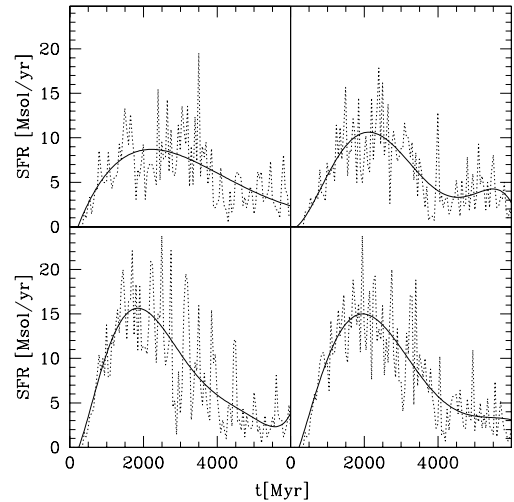}
\caption{Star formation histories. Top left panel: model A; top
right: model B; bottom left: model C;  bottom right: model D. See
text for details.} \label{sfr_toy_comp}
\end{figure}

\begin{figure}
\centering
\includegraphics[width=7cm,height=7cm]{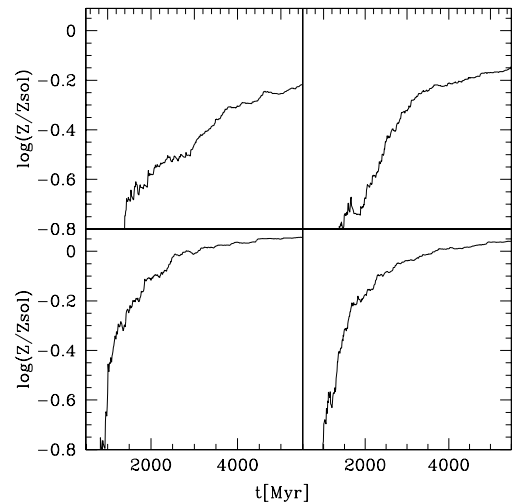}
\caption{Chemical enrichment histories. Top left panel: model A; top
right: model B; bottom left: model C;  bottom right: model D. See
text for details.} \label{chemo_toy_comp}
\end{figure}

\textbf{Star formation and energy feedback}. The relationship
between star formation rate and energy feed back deserves particular
attention and some more detailed investigation. In general, one
expects that the higher the energy given to the interstellar medium
by SN-explosions, UV radiation, stellar winds, the lower is the
efficiency of star formation. This holds  in particular with our
scheme in which only cold gas is allowed to form stars. Our
reference model (the one we have presented) contains four sources of
energy feedback: SN-explosions, UV radiation, stellar winds, and a
contribution of radial kinetic energy imparted to gas clouds
surrounding a dying star particle suffering SN-explosions (see Sec.
\ref{galdyn}, point 6). This simple scheme could be too a crude
description of reality. In addition to that, our detailed history of
star formation proceeds in a number of short bursts of activity,
which could be ascribed partly to the scheme we have adopted to deal
with energy feedback and partly to numerical resolution (this
latter point will be examined below). In order to investigate  the
effect of energy feedback on the star formation rate, we run four
simulations with the same initial conditions and total mass of the
galaxy model we have being discussing in so far. These four models
are, however, calculated with 3500 particles of baryonic mass and
3500 particles of dark matter, i.e. half of the number for each
component contained in the original model. The reason of this choice
will become clear when discussing the effect of the numerical
resolution. The difference among the four models is only in the
energy feedback: (1) Model A is the same as the original model but
for the fewer number of particles. It contains all sources of energy
feedback. (2) Model B contains only the SN-explosions, UV radiation
and stellar winds (no contribution by radial kinetic energy to
nearby gas particle). (3) Model C contains only the SN-explosion and
contribution of kinetic energy to nearby gas particles (no UV and
stellar winds inputs are considered). (4) Finally, model D has only
the contribution from SN-explosions, while all the rest is quenched
off. Fig. \ref{sfr_toy_comp} compares the results for star formation
(as in Fig. \ref{sfr_comp}, in this case too polynomial best fits of
the star formation rate are superposed to the fluctuating star
formation rate of the numerical model to highlight the general
trends). Fig. \ref{chemo_toy_comp} compares the results for chemical
enrichment in the four models. As expected, the effects are indeed
remarkable. Including more and more sources of energy feedback
tends to lower the mean star formation activity and to quench it
much earlier as compared to the cases in which they are partially
included. Consequently, chemical enrichment also stops and saturates
much earlier.

\begin{figure}
\centering
\includegraphics[width=7cm,height=7cm]{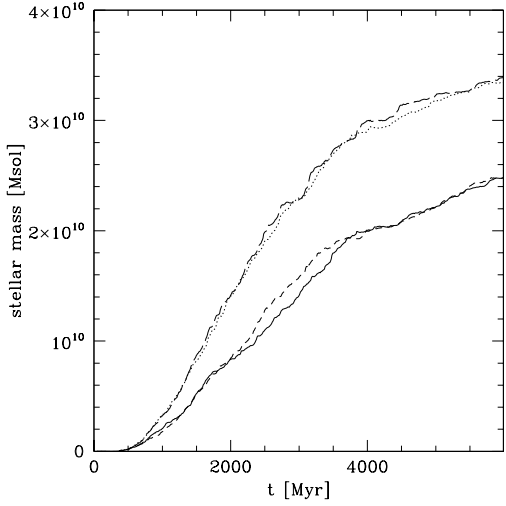}
\caption{Cumulative star formation histories for models
A (solid), B (short dashed), C (long dashed) and D
(dotted - see text for details).}
\label{cumul}
\end{figure}

The separated contribution by the different sources of
energy feedback is best illustrated in Fig. \ref{cumul} showing the
cumulative building up of the stellar mass for the four models.
Models A and B contain all thermal energy sources (SN, UV, Winds)
and differ for the kinetic kick. The same applies  to the pair C and
D, which however contains only the SN input for the thermal part.
Little or no difference is caused by the kinetic energy kick
(compare  A with B and  C with  D). The largest difference in the
cumulative stellar mass is due to the differences in the thermal
energy input: star formation efficiency and hence total stellar mass
decrease at increasing feedback. 

The star formation history always shows a high number
of burst episodes, whose amplitude seems to increase at decreasing
the number of energy feedback terms. However, this can be partly
ascribed to the lower number of particles used in these simulations,
as discussed next.

\begin{figure}
\centering
\includegraphics[width=7cm,height=7cm]{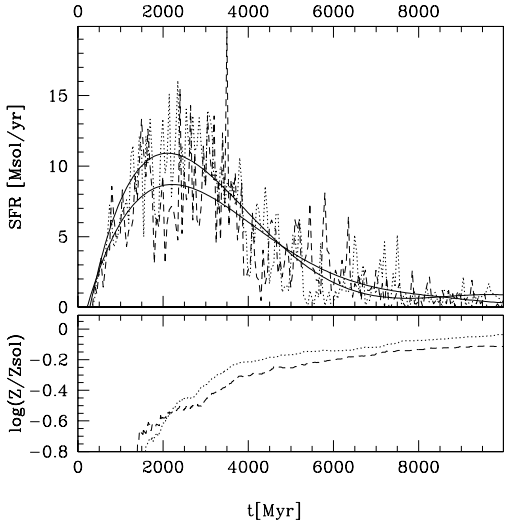}
\caption{Comparison of the star formation histories (top panel) for
the original model with 14,000 particles (about 7000 of Dark Matter
and 7000 of Baryonic Matters) and full energy input indicated by the
long dashed line  and model A with the same energy input but 7000
particles in total (about 3500 of Dark Matter and 3500 of Baryonic
Matter) indicated by the dotted line. The polynomial fits to star
formation rate are shown for both cases to guide the eye in the
gross evaluation of the star formation intensity and its temporal
evolution. The bottom panel show the same but for the chemical
enrichment history } \label{sfr_chemo_comp}
\end{figure}

\textbf{Numerical resolution and physical results}. What is
the effect of the number of particles (mass resolution) used in our
models on the physical consistency of the results? This is an
important issue to address considering that multi-phase NB-TSPH
simulations of galaxy formation and evolution with significantly
higher numbers of particles can be found in literature: for instance
\citet[][ from 20,000 to 250,000 particles]{Booth07},  \citet[][
from 18,000 to 80,000 particles]{Scannapieco2006}, and \citet[][
35,000 particles]{Harfst06}. These studies present also higher
number of cases, whereas we present only one case. Therefore,
compared to those studies our resolution and number of models are
somewhat limited. As far as the number of models is concerned we put
it aside because a statistical analysis of the variety of possible
results (model galaxies) is beyond the aims of the present study.
The crucial question to address is whether our resolution is
sufficient to yield physically consistent results. We have already
mentioned that a number of studies have been made based on a rather
small number of particles comparable to ours, e.g. \citet[][ 10000
of dark Matter and 10000 of Baryonic Matter]{Buonomo00}, \citet[][
who lowered the particles of Baryonic Matter down to a few
thousands]{Buonomo99} and \citet[][ who used 50000 particles of
Baryonic Matter and 10000 of Dark Matter]{Semelin02}.  All these
models adopted "isolated initial conditions". Other studies adopt
cosmological initial conditions and numbers of particles comparable to ours
\citep[see e.g.][]{Kawata99, Kawata01a,Kawata01b,Kawata03}.

\textsl{Star Formation}. It is worth comparing the star
formation and chemical enrichment histories of our fiducial model
with those of case A. The models have exactly the same initial conditions,
total mass, and energy input, but they differ in the number of
particles (14000 versus 7000 in total). This is shown in the top and
bottom panels of Fig. \ref{sfr_chemo_comp}. In the top panel of the
star formation rate we also plot the polynomial best fits to guide
the visual inspection of the results. The overall trend is the same,
even if the star formation rate of model A is more fluctuating and
slightly less intense than that of the original model. The difference at the
activity peak is $\sim$ 23\%, which is immediately mirrored by the
chemical enrichment history (bottom panel) where the maximum
metallicity is $\sim$ 15\% less than in the original model. These
effects are clearly caused by the number of particles. As far as the
fluctuations in the star formation rate are concerned, there are at
least three causes: (i) the stochastic treatment of star formation
rate and energy input by SN explosions, stellar winds and UV
radiation; (ii) the small number of particles (in particular those
prone to star formation) and (iii) the age sampling of the numerical
results, which are displayed averaging the numerical results over
time intervals of 50 Myr. Certainly, by increasing the number of
particles and in particular those of cold molecular gas (see below),
they should smoothen out, if not disappear. Therefore we dot
attribute them any particular physical meaning. What matters here if
the global trend represented by the analytical fits. One may guess
that increasing the number of particles to about 50000 would be the
ideal situation. The parallel version of the same code is in the
final testing phase (at CINECA Super-Computing Centre), so that
simulations with much higher numbers of particles are next in our
plans.

\textsl{Number of molecular clouds}. At each time step, the
number of cold gas particles in the fiducial model is of order of one hundred,
which leads to a strong Poisson noise for star formation rate, which is enhanced
by cloud-cloud collisions. Does it reflect onto the structural
properties, e.g. the Sersic-like mass density profile of our model
galaxy? In Section 3.3 we have provided a plausible explanation why
in the physical situation of our models aimed at simulating
elliptical galaxies the number of cold clouds remains small as
compared to that expected in a disc galaxy. It will not be repeated
here. In any case, the number of cold gas particles is
self-consistent with the physical input of the models. Increasing it
further is not possible. Whether increasing the number of particles,
we would end up with a disc galaxy instead of an elliptical one is
the subject of the next paragraph.

\textsl{Disc or elliptical galaxies?} The models we have calculated
(the fiducial one in particular) strictly resemble an elliptical
galaxy (as far as we can tell looking at the structural parameters,
surface mass density profile and axial ratios).  Can a simulation,
calculated with a larger number of particles but the same initial
conditions and physical input, end up as disc galaxy?  We do not
think so. Very little angular momentum is initially given to the
proto-galaxy, and there are no external torques which could induce
sizeable rotational effects. Therefore, the formation of an almost
spherical structure is  a straight consequence. It has a reasonable
density profile and a stable relaxed configuration which resembles
real spherical/elliptical galaxies. Likely perturbations with little
initial rotation and weak external torques  will end up with a
nearly spherical shape and  Sersic profile of the mass distribution.

Finally, similar results have
been presented by \citep[see e.g.][]{Kawata99,
Kawata01a,Kawata01b,Kawata03} using similar initial conditions,
number of particles, and input physics, but for the multi-phase
description of interstellar medium and star formation efficiency.

\section{Discussion and conclusions} \label{disc}

The new description of the ISM allows us to get a deeper insight on
the time scales and processes of star formation, even at the
relatively low resolution of the present model. Good agreement with
important observational information is achieved, especially for what
concerns the  mass assembly time scales and chemical evolution of
the system. The agreement can  be further improved by finely tuning
some of the model parameters.

More and more observational evidences hint the existence of a
population of big, red, and isolated galaxies already in place up to
z $\sim 6$ \citep[see e.g.][]{Mobasher05, Cimatti99, Cimatti04}.
Moreover, recent large surveys like DEEP2, SDSS etc. clearly
demonstrate that not only the stellar populations of massive
galaxies are older than the ones belonging to smaller systems
\citep[\textit{downsizing}, ][]{Bundy05}\footnote{The
concept of \textit{downsizing} star formation was first suggested by
 \citet{Matteucci94} from variations of the [Fe/Mg]
 ratio with the galaxy luminosity, by \citet{Bressan96}  from the analysis
 of the line absorption indices and UV excess, and by \citet{Cowie96}.}
 but that \textit{massive galaxies assembled their mass at higher
redshift than the less massive ones} \citep[\textit{top-down}
formation, ][]{Bundy06,Cimatti06}. It is soon clear that both
\textit{downsizing} and \textit{top-down} views of star formation
and mass assembly are in conflict with the "classical" predictions
of the hierarchical theories of structure formation, whereas they
can be easily framed into the monolithic (as explained above)
view of galaxy formation.
Despite the many attempts that are being made to
reconcile the hierarchical scenario with the new compelling
evidences of "very early and nearly monolithic" assembly of early
type galaxies - \textit{dry mergers} \citep{Bell04} and
\textit{hierarchical down-sizing} \citep{deLucia06} are the most
popular ones - the observational data leave little room for viable
solutions in the context of the hierarchical scheme. In particular,
we note that dry mergers cannot have played a fundamental role in
the formation of early-type galaxies, given the weak dependence of
their formation time-scales on environmental conditions and  weak
cosmic evolution of the  mass function of galaxies \citep{Bundy06}.
Finally, \citet{Clemens06} show how a burst of monolithic star
formation in high mass galaxies and a slower, less efficient,
ongoing star formation activity in smaller objects are necessary
conditions to achieve the observed trends of metallicity and
$\alpha$-enhancement. Remarkably, \textit{down-sizing} and
\textit{top-down} are natural products of the models presented by
\citet{ChiosiCarraro02} (see they star formation histories as a
function of the total mass and initial density).

A monolithic-like star formation as the one we present here \citep[as
well as in previous studies, see e.g.][]{Merlin06}, with more or
less prolonged durations of star formation activity (depending on
the total mass and initial density of the proto-galaxy, and perhaps
on the environment in which it is being formed) seems to be able to
match most of the characteristics of old, massive spheroids, as well
as of younger and bluer objects. We plan to perform an extended
study on the dependence of the star formation history on the mass
and density of the  proto-galaxies, using the parallel version of
\textsc{GalDyn} in which the new prescription is being implemented. The
longer period of activity found in the new models agrees with the
trends suggested by the observational data. Fine tuning of the model
parameters is the first step to undertake to reproduce the pattern
of  observational constrains. The goal is to account for key
diagnostic planes such as the classical Fundamental Plane for
local ellipticals \citep{Djorg87}, the \textit{mass quenching}
relation found by \citet{Bundy06}, and the  chemical trends found by
\citet{Gallazzi05}.

Given this picture, it is no longer necessary to consider mergers
between proto-galaxies (or disks) as the main way in which massive
ellipticals are formed. It is most likely, indeed,  that this
scenario has to be definitely discarded. If so, a general scenario
reconciling our understanding  of matter aggregation in the Universe
on large scales with the compelling evidences about the formation of
galaxies is missing. Likely, Nature follows the hierarchical mode
when aggregating matter on the scale of groups and clusters, and the
monolithic-like mode when aggregating (baryonic) matter on the scale
of individual galaxies. On the other hand galaxy mergers cannot be
completely ruled out, simply because we have direct observational
evidence of this phenomenon. They are beautiful, spectacular events,
but not the dominant mechanism by which galaxies (the early-type in
particular) are assembled and their main features
imprinted. In the forest of galaxy formation theories, \textit{ex
pluribus unum.}

\begin{acknowledgements}
The authors would like to thank Prof. Guido Barbaro for the helpful
suggestions. We would also thank the anonymous referee  for her/his
critic and very constructive remarks that much improved the final
version of this study. Spurred indeed by her/his questions we have
addressed and commented aspects of the problem that were not
initially taken into consideration. This study was financed by the
University of Padua by supporting the PhD fellowship assigned to E.
Merlin, the Department of Astronomy by indirectly providing some
facilities (computers, infrastructure, etc), and  the
Super-Computing Center CINECA (Bologna, Italy) by allocating  a
substantial amount of computing time. No financial resources have
been obtained from other National Agencies (MIUR).
\end{acknowledgements}


\bibliographystyle{apj}           

\end{document}